\documentclass[aps,pra,twocolumn,floatfix,nofootinbib]{revtex4-2}
\usepackage[colorlinks]{hyperref}
\usepackage[utf8]{inputenc}
\usepackage{graphicx}
\usepackage{tabularx}
\usepackage{multirow}
\usepackage{amssymb}
\usepackage{amsmath}
\usepackage{microtype}
\usepackage{xcolor}
\usepackage{dcolumn}
\usepackage{refcount}
\usepackage{natbib}
\DeclareMathAlphabet\mathbfcal{OMS}{cmsy}{b}{n}

\newcommand{\bra}[1]{\ensuremath{\langle #1|}}	%Dirac Bras
\newcommand{\ket}[1]{\ensuremath{|#1\rangle}}	%Dirac Kets
\newcommand{\threej}[6]{\ensuremath{\begin{pmatrix}#1&#2&#3\\#4&#5&#6\end{pmatrix}}}	% 3-j symbol
\newcommand{\sixj}[6]{\ensuremath{\begin{Bmatrix}#1&#2&#3\\#4&#5&#6\end{Bmatrix}}}	% 6-j symbol

\renewcommand{\v}[1]{\ensuremath{\boldsymbol{#1}}}		%bold-math for vectors
\newcommand{\brad}[1]{\ensuremath{\langle #1||}}
\newcommand{\ketd}[1]{\ensuremath{|| #1\rangle}}
\makeatletter
\newcolumntype{d}[1]{D{.}{.}{#1}}
\newcommand{\doublewidetilde}[1]{{%
  \mathpalette\double@widetilde{#1}%
}}
\newcommand{\double@widetilde}[2]{%
  \sbox\z@{$\m@th#1\widetilde{#2}$}%
  \ht\z@=.9\ht\z@
  \widetilde{\box\z@}%
}

\makeatother 
\begin{document}

\begin{abstract}
Extracting electroweak observables from experiments on atomic parity violation (APV) using the Stark interference technique requires accurate knowledge of transition polarizabilities.  In cesium, the focus of our paper, the $6S_{1/2}\rightarrow{7S_{1/2}}$ APV amplitude is deduced from the measured ratio of the APV amplitude to the vector transition polarizability, $\beta$. This ratio was  measured with a $0.35\%$ uncertainty by the Boulder group [Science {\bf 275}, 1759 (1997)]. Currently, there is a sizable discrepancy in different determinations of $\beta$ critically limiting the interpretation of the APV measurement. The most recent value [Phys.\ Rev.\ Lett.\ {\bf 123}, 073002 (2019)] of $\beta=27.139(42)\, \mathrm{a.u.}$ was deduced from a semi-empirical sum-over-state determination of the scalar transition polarizability $\alpha$ and the measured $\alpha/\beta$ ratio [Phys.\ Rev.\ A {\bf{55}}, 1007 (1997)]. This value of $\beta$, however, differs by $\sim 0.7\%$ or $2.8\sigma$ from the previous determination of $\beta=26.957(51)$ by [Phys.\ Rev.\ A {\bf{62}}, 052101 (2000)] based on the measured ratio $M1/\beta$ of the magnetic-dipole $6S_{1/2}\rightarrow{7S_{1/2}}$ matrix element to $\beta$. Here, we revise the determination of $\beta$ by [Phys.\ Rev.\ Lett.\ {\bf 123}, 073002 (2019)], using a more consistent and more theoretically complete treatment of contributions from the excited intermediate states in the sum-over-state $\alpha/\beta$ method. Our result of  $\beta=26.887(38)\, \mathrm{a.u.}$ resolves the tension between the $\alpha/\beta$ and $M1/\beta$ approaches. We recommend the value of $\beta=26.912(30)$ obtained by averaging our result and that of [Phys.\ Rev.\ A {\bf{62}}, 052101 (2000)].
\end{abstract}

\title{Reevaluation of Stark-induced transition polarizabilities in cesium}
\author{H. B. Tran Tan}
\author{D. Xiao}
\author{A. Derevianko}
\email[]{andrei@unr.edu}
\affiliation{Department of Physics, University of Nevada, Reno, 89557, USA}
\date{\today}
\maketitle

\section{Introduction}

Atomic parity violation (APV) plays an important role in probing the electroweak sector of the standard model (SM) of elementary particles at low energy. The information derived from table-top APV experiments is both  
complementary to and in competition with that from large-scale particle colliders (see, e.g., the review~\cite{SafBudDeM2018.RMP} and references therein). To date, the 1997 Boulder experiment~\cite{WooBenCho97} searching for APV in $^{133}\mathrm{Cs}$ remains the most accurate. A substantial body of work has been devoted to the interpretation of and the extraction of electroweak observables from the Boulder results. 

In its setup, the Boulder experiment~\cite{WooBenCho97} employed the ${}^{133}$Cs $6S_{1/2}\rightarrow{7S_{1/2}}$ transition, whose $E1$ amplitude nominally vanishes due to the parity selection rule. However, parity nonconserving (PNC) weak interactions between the atomic nucleus and electrons
admix small components of $P_{1/2}$ states into the nominal $S_{1/2}$ states, thus opening the $E1$ channel.
Using the parity-mixed multi-electron states $\ket{6S'_{1/2}}$ and $\ket{7S'_{1/2}}$ and the hyperfine basis (see Eq.~\eqref{eq:hyperfine-basis} below), the APV transition amplitude may be written as
\begin{align}\label{eq:E1_PNC1}
    A^{\mathrm{PNC}}_{fi}&=\bra{7S_{1/2}',\,F_fM_f}-\mathbfcal{E}_L\cdot{\v{D}}\ket{6S_{1/2}',\,F_iM_i} \nonumber\\
    &=i\mathrm{Im}(E1_{\mathrm{PNC}})\mathbfcal{E}_L\cdot\bra{F_fM_{F_f}}\v{\sigma}\ket{F_iM_{F_i}}\,,
\end{align}
where $\mathbfcal{E}_{L}$ is the laser electric field driving the $E1$ transition, $\v{D}$ is the electric dipole operator, and $\v{\sigma}$ is the Pauli matrix.

Due to the smallness of $E1_{\rm{PNC}}$, which is on the order of $\sim10^{-11}$ in atomic units, measuring the PNC transition amplitude $A^{\mathrm{PNC}}_{fi}$ directly is a formidable challenge. To overcome this difficulty, it was suggested that one uses the Stark-interference technique~\cite{BouBou74,BouBou75}, which relies on the mixing of states of opposite 
parities due to an externally applied electric field. The transition rate $R$ between the parity-mixed states then includes contributions from the Stark-induced $E1$, magnetic-dipole $M1$, and the PNC-induced amplitudes~\cite{WooBenRob99}
\begin{equation}
R\propto|A^{\mathrm{Stark}}_{fi}+A^{M1}_{fi}+A^{\mathrm{PNC}}_{fi}|^2\,.
    \label{eq:rate-equation}
\end{equation}
Upon expansion, the right-hand side of Eq.~\eqref{eq:rate-equation} yields the Stark-PNC interference term, $2\mathrm{Re}[A^{\rm{Stark}}_{fi}(A^{\rm{PNC}}_{fi})^{*}]$, whose sign is subject to the handedness of the experiment measuring $R$. Thus, the PNC amplitude $A^{\mathrm{PNC}}_{fi}$ can be extracted from the Stark-PNC interference term by measuring the changes in $R$ under parity reversals. Based on the Stark-interference technique, the Boulder group~\cite{WooBenRob99} reported the following values
\begin{equation}\label{Eq:BoulderResults} 
    \frac{\mathrm{Im}(E1_{\mathrm{PNC}})}{\beta}=\begin{cases}
    -1.6349(80)\, \mathrm{mV/cm}\\
    \,\,\,\,{\rm for}\,\,\,6S_{1/2},\,{F_i=4}\rightarrow{7S_{1/2},\,{F_f=3}}\,,\\
    -1.5576(77) \, \mathrm{mV/cm} \\
    \,\,\,\,{\rm for}\,\,\,6S_{1/2},\,{F_i=3}\rightarrow{7S_{1/2},\,{F_f=4}}\,,
    \end{cases}
\end{equation}
where $\beta$ is the atomic vector polarizability. 

A weighted average of the two values in Eq.~\eqref{Eq:BoulderResults} yields the nuclear-spin-independent observable, i.e., the nuclear weak charge, while their difference determines nuclear-spin-dependent effects, e.g., the nuclear anapole moment. For the extraction of these quantities, knowledge of the vector transition polarizability $\beta$ is essential and substantial attention~\cite{Dzuba1997,Safronova1999,Vas2002,Dzuba2002,Toh2019,Ben1999,DzuFla00} has been paid over the years to determining its value. Since 2000, the most accurate value of $\beta$ has been determined based upon combining a semi-empirical calculation of the hyperfine-induced magnetic-dipole $6S_{1/2}\rightarrow7S_{1/2}$ transition amplitude $M1$~\cite{DzuFla00} with a measurement of the ratio $M1/\beta$~\cite{Ben1999}. Another approach to estimating $\beta$ combines a calculation of the scalar polarizability $\alpha$ with the measurement of the ratio $\alpha/\beta$~\cite{ChoWooBen97}. The latest most accurate determination of $\beta$ was published by the Purdue group~\cite{Toh2019} who adopted the most accurate value of $\alpha/\beta=9.905(11)$~\cite{ChoWooBen97} and used the
sum-over-state (SoS) method to calculate $\alpha$. Their calculation of $\alpha$ were carried out using experimentally and theoretically determined matrix elements and energies. Although the uncertainties of the $\alpha/\beta$~\cite{ChoWooBen97,Toh2019} and $M1/\beta$~\cite{Ben1999,DzuFla00} approaches are comparable, both approximately at the level of $0.2\%$, their central values differ by $\sim 0.7\%$ or 2.7$\sigma$. This difference critically undermines the accuracy of extracting electroweak observables from the Boulder APV measurement.

\begin{figure}[!ht]
    \centering
    \includegraphics[width=\columnwidth]{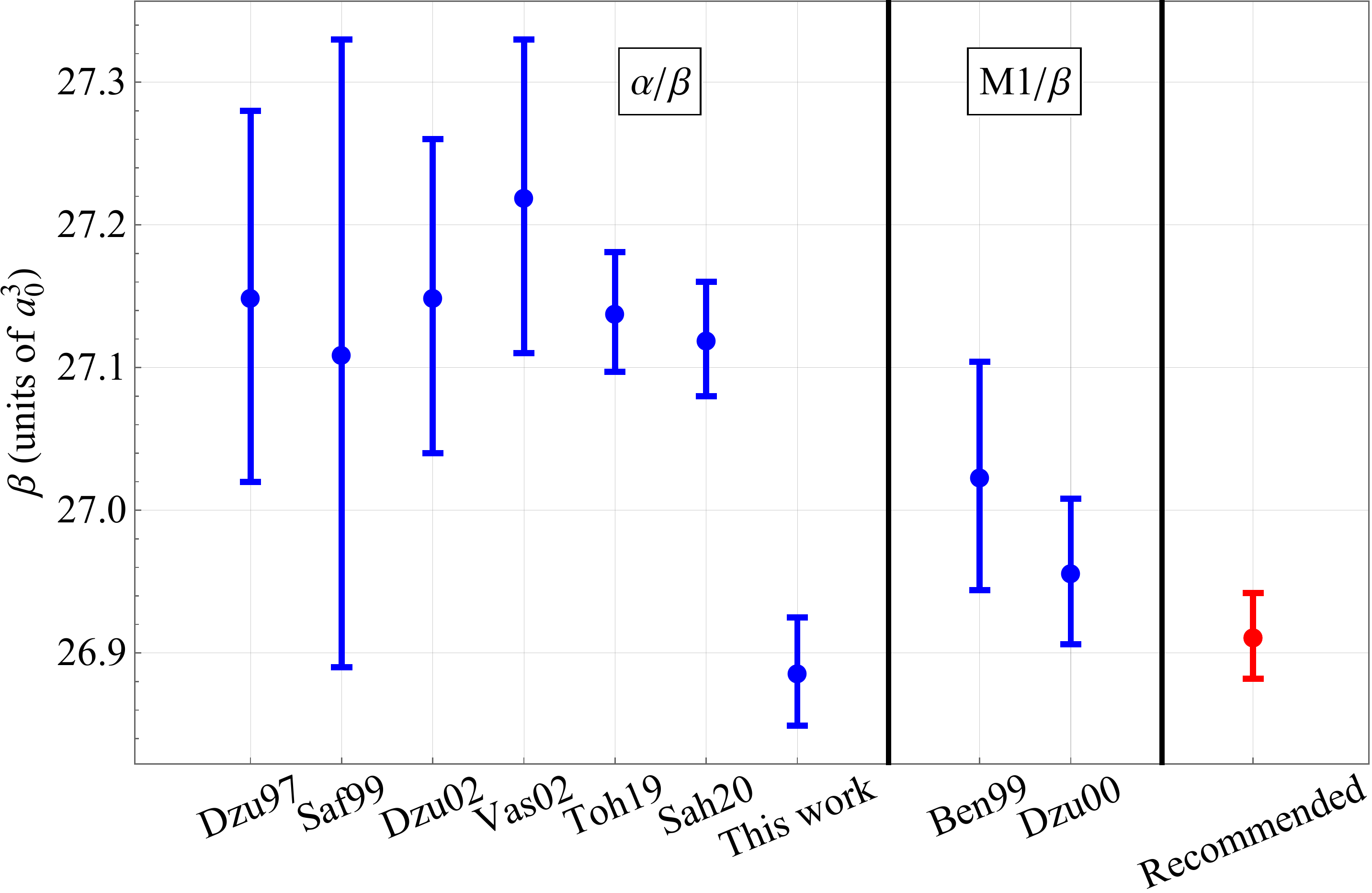}
    \caption{Comparison of our value for the vector transition polarizability $\beta$ with previous results~\cite{Dzuba1997,Safronova1999,Vas2002,Dzuba2002,Toh2019,Ben1999,DzuFla00}. The previous determinations of $\beta$ are identified by the initial three letters of the first author's last name and the abbreviated publication year. The left panel presents results from the sum-over-state approach, the middle panel those from the $M1/\beta$ determination, and the right panel shows our recommended value for $\beta$ obtained by taking a weighted average of our result and that of Ref.~\cite{DzuFla00}.}\label{Fig:beta}
\end{figure}

Recently, our theory group performed the most sophisticated to date \textit{ab initio} calculations of the $E1$ transition matrix elements in Cs~\cite{tan2023precision}. Here, we use these newly determined $E1$ matrix elements and a SoS approach to reevaluate the scalar and vector polarizabilities $\alpha$ and $\beta$. We show that the updated value of $\beta$ agrees well with that obtained from the $M1/\beta$ method of Refs.~\cite{Ben1999,DzuFla00}, thus reconciling the two alternative approaches (see Fig.~\ref{Fig:beta}).

The paper is organized as follows. In Sec.~{\ref{Sec:Review-Stark-induced-transition}}, we provide a review and derivation for the second-order transition polarizabilities $\alpha$ and $\beta$. In Sec.~{\ref{Sec:Cal-sec-order}}, we detail the numerical methods employed for the computation of these quantities. In Sec.~\ref{Results and error estimates}, we present our numerical values and error estimates and provide a comparison with previous results. Unless stated otherwise, atomic units are used throughout.

\section{Stark-induced \texorpdfstring{$E1$}{} transitions and transition polarizabilities}\label{Sec:Review-Stark-induced-transition}

In the presence of a DC electric field, the initial and final $S$ states of Cs admix states of opposite parities, thus enabling the otherwise forbidden $E1$ transition between the $6S_{1/2}$ and $7S_{1/2}$ states. In this section, we rederive the conventional results for the Stark-induced $E1$ transition amplitudes. The reader may refer to the original paper by~\citet{BouBou75} for an alternative derivation.

We start by introducing the hyperfine basis
\begin{equation}\label{eq:hyperfine-basis}
    \ket{n\,({IJ})FM_F}=\sum\limits_{M_JM_I}C^{FM_F}_{JM_JIM_I}\ket{n\,JM_J}\ket{IM_I}\,,
\end{equation}
whose members are formed by coupling electronic states $\ket{n\,JM_J}$ of angular momentum $\v{J}$ and nuclear states $\ket{IM_I}$ of spin $\v{I}$ to form states of definite total angular momentum $\v{F}=\v{I}+\v{J}$. Here, $M_F$, $M_J$, and $M_I$ are the magnetic quantum numbers, $n$ stands for the remaining quantum numbers, such as the principal quantum number of the electronic state, and $C^{FM_F}_{JM_JIM_I}$ is the conventional Clebsch-Gordan coefficients. %The state $\ket{n\,({IJ})FM_F}$ has energy $E_{nJ}+E_I$ where $E_{nJ}$ is the electronic energy and $E_I$ is the nuclear energy.

In the hyperfine basis, Eq.~({\ref{eq:hyperfine-basis}}), the initial and final states involved in the $i\rightarrow f$ transition are
\begin{subequations}\label{eq:zeroth-order}
    \begin{align}  
        \ket{i}&\equiv\ket{n_i(IJ_i)F_iM_i}\,,\\
        \ket{f}&\equiv\ket{n_f(IJ_f)F_fM_f}\,.
    \end{align}
\end{subequations}
In the presence of an externally applied static electric field $\mathbfcal{E}_{S}$, the initial and final states acquire the admixtures
\begin{subequations}
    \begin{align}
        \ket{\delta i}&=-\sum_{a\neq i} \ket{a}\frac{{\mathbfcal{E}}_S \cdot \v{D}_{ai}}{\Delta{E_{ia}}}\,,\label{eq:6a}\\
        \ket{\delta f}&=-\sum_{a\neq f}\ket{a}\frac{{\mathbfcal{E}}_S\cdot\v{D}_{af}}{\Delta{E_{fa}}}\,,\label{eq:6b}
    \end{align}
\end{subequations}
where $\Delta{E_{ab}}\equiv{E_a-E_b}$ and $\v{D}_{ab}\equiv \bra{a}\v{D}\ket{b}$ is the electric dipole matrix element.

If a laser is now applied, it can drive the transition $i\rightarrow{f}$, whose Stark-induced $E1$ transition amplitude is given by
\begin{align}\label{eq:stark-induced-amp}
    A_{fi}&=-\bra{\delta f}\v{{\mathbfcal{E}}}_L\cdot{\v{D}}\ket{i}-\bra{f}\v{{\mathbfcal{E}}}_L\cdot{\v{D}}\ket{\delta i}={{\mathcal{E}}}_L{{\mathcal{E}}}_S a_{fi}\,,
\end{align}
where $\mathbfcal{E}_L$ is the laser electric field. In the last step of Eq.~\eqref{eq:stark-induced-amp}, we have factored out the amplitudes of the electric fields and defined the Stark-induced transition polarizability
\begin{align}\label{eq:st_amp_exp0}
    a_{fi}&\equiv\sum_{a\neq f}\frac{(\v{\hat{\varepsilon}}\cdot{\v{D}}_{fa})(\v{\hat{e}}\cdot{\v{D}}_{ai})}{\Delta{E_{fa}}}\nonumber\\
    &+\sum_{a\neq i}\frac{(\v{\hat{e}}\cdot{\v{D}}_{fa})(\v{\hat{\varepsilon}}\cdot{\v{D}}_{ai})}{\Delta{E_{ia}}}\,.
\end{align}
Note that $a_{fi}$ still depends on the polarization vectors $\hat{\v{e}}\equiv\v{\mathcal{E}}_S/\mathcal{E}_S$ and $\hat{\v{\varepsilon}}\equiv\v{\mathcal{E}}_L/\mathcal{E}_L$ of the DC and laser fields.

The expression for $a_{fi}$ may be cast into a form more convenient for angular reduction. To this end, one uses the recoupling identity~\cite{VarMosKhe88}
\begin{align}\label{eq:identities}
    (R^{(k_1)}\cdot S^{(k_1)})(U^{(k_2)}\cdot{V^{(k_2)}}) &= \sum\limits_{Q}(-1)^{Q-k_1-k_2}\nonumber\\
    \times\{R^{(k_1)}\otimes U^{(k_2)}\}^{(Q)}&\cdot{\{S^{(k_1)}\otimes V^{(k_2)}}\}^{(Q)}\,,
\end{align}
where the operators ${P^{(k_1)}}$, ${Q}^{(k_1)}$, ${R}^{(k_2)}$, and ${S}^{(k_2)}$ are irreducible tensor operators (ITOs) of ranks $k_1$ and $k_2$. In Eq.~\eqref{eq:identities}, a scalar product of two rank-$k$ ITOs is understood as the following sum over their spherical components
\begin{equation}
    {P^{(k)}}\cdot{{Q}^{(k)}} = \sum_{q=-k}^{k} (-1)^q P^{(k)}_{q} {Q}^{(k)}_{-q}\,,
\end{equation}
and a compound ITO of rank $Q$ is defined as 
\begin{equation}
\{P^{(k_1)}\otimes{R^{(k_2)}}\}_{q}^{(Q)}=\sum_{q_1q_2} C^{Qq}_{k_1q_1k_2q_2} P^{(k_1)}_{q_1} R^{(k_2)}_{q_2}\,,
\end{equation}
where $q_1$ and $q_2$ label the spherical basis components of the ITOs.
The possible values of $Q$ are limited by the triangular selection rule, i.e., ${|k_1-k_2|}\le{Q}\le{k_1+k_2}$.

In our case of the electric dipole couplings, the polarization and dipole operators in Eq.~({\ref{eq:identities}}) are ITOs of rank $1$. As a result, one has
\begin{align}\label{eq:st_amp_exp}
a_{fi}&=\sum\limits_{Q=0}^2(-1)^Q\{\hat{\v{\varepsilon}}\otimes\hat{\v{e}}\}^{(Q)}\cdot\left(\sum_{a\neq f}\frac{\{\v{D}_{fa}\otimes{{\v{D}_{ai}}}\}^{(Q)}}{\Delta{E_{fa}}} \right. \nonumber\\
&\left. +(-1)^{Q}\cdot\sum_{a\neq i}\frac{\{\v{D}_{fa}\otimes{\v{D}_{ai}}\}^{(Q)}}{\Delta{E_{ia}}} \right)\,,
\end{align}
where we have used $\{\hat{\v{\varepsilon}}\otimes\hat{\v{e}}\}_{q}^{(Q)}=(-1)^Q\{\hat{\v{e}}\otimes\hat{\v{\varepsilon}}\}_{q}^{(Q)}$. The term in Eq.~\eqref{eq:st_amp_exp} with $Q=0$ corresponds to the scalar, that with $Q=1$ to the vector, and the one with $Q=2$ to the tensor (quadrupole) contributions to the transition polarizability.  

To simplify Eq.~\eqref{eq:st_amp_exp} further, one may introduce the effective ITOs,
\begin{equation}\label{Eq:aITOs}
   a[k]_{q}^{(Q)}\equiv\{\v{D}\otimes{R_k}\v{D}\}^{(Q)}_{q}\,,
\end{equation}
with the resolvent operator $R_k\equiv{(E_k-H_0)^{-1}}$, where $H_0$ stands for the unperturbed atomic Hamiltonian. Since $R_k$ has the spectral resolution
\begin{align}\label{eq:resolvent-op}
R_k=(E_k-H_0)^{-1}=\sum\limits_{a\ne{k}}\Delta{E_{ka}^{-1}}\ket{a}\bra{a}\,,
\end{align}
and is a scalar (so that the combination $R_k\v{D}$ remains a rank-1 ITO), Eq.~({\ref{eq:st_amp_exp}}) may be written as
\begin{align}\label{eq:st_amp_exp_redux}
a_{fi}&=\sum\limits_{Q=0}^2(-1)^Q\{\hat{\v{\varepsilon}}\otimes\hat{\v{e}}\}^{(Q)}\nonumber\\
&\cdot\left(\bra{f}a[f]^{(Q)}\ket{i}+(-1)^{Q}\bra{f}a[i]^{(Q)}\ket{i} \right)\,.
\end{align}
which, upon applying the Wigner-Eckart theorem, further simplifies to
\begin{align}\label{eq:st_amp_simple_exp}
a_{fi}&=\sum\limits_{Q=0}^{2} w_Q( \hat{\v{\varepsilon}},\hat{\v{e}}) \nonumber\\
&\times\left[\brad{f}a[f]^{(Q)}\ketd{i}+(-1)^{Q}\brad{f}a[i]^{(Q)}\ketd{i}\right]\,,
\end{align}
where the multipolar polarization weights $w_Q( \hat{\v{\varepsilon}},\hat{\v{e}})$ are defined as
\begin{align}\label{eq:w_q}
w_Q( \hat{\v{\varepsilon}},\hat{\v{e}})&=(-1)^{Q}\sum_q(-1)^{q+F_f-M_f}\nonumber\\
&\times\threej{F_f}{Q}{F_i}{-M_f}{-q}{M_i}(\hat{\v{\varepsilon}}\otimes{\hat{\v{e}}})_{q}^{(Q)}\,,
\end{align}
where $\threej{F_f}{Q}{F_i}{-M_f}{-q}{M_i}$ is the $3j$ symbol.

Finally, by summing over magnetic quantum numbers, we obtain
\begin{align}\label{eq:afq_exp}
\brad{f}a[f]^{(Q)}\ketd{i}&=(-1)^{F_i+I-J_i}[F_f,Q,F_i]^{1/2}\nonumber\\
&\times\sixj{Q}{J_f}{J_i}{I}{F_i}{F_f}\sum\limits_{n_aJ_a}\sixj{Q}{J_i}{J_f}{J_a}{1}{1}\nonumber\\
&\times\frac{\brad{n_fJ_f}D\ketd{n_aJ_a}\brad{n_aJ_a}D\ketd{n_iJ_i}}{E_{n_fJ_f}-E_{n_aJ_a}}\,,
\end{align}
where $[J_1,J_2,\ldots,J_n]\equiv(2J_1+1)(2J_2+1)\ldots(2J_n+1)$.
The reduced matrix elements $\brad{f}a[i]^{(Q)}\ketd{i}$ are given by the same formula, but with $E_{n_iJ_i}$ replacing $E_{n_fJ_f}$ in the energy denominator. We point out that due to the $6j$ symbols in Eq.~\eqref{eq:afq_exp}, the term with $Q=2$ vanishes for $J_{f}=J_{i}=1/2$, in particular for the transition $6S_{1/2}\rightarrow{7S}_{1/2}$ of interest, as expected.

Conventionally, the Stark-induced transition polarizability $a_{fi}$ is expressed as a linear combination of the second-order scalar and vector polarizabilities~\cite{BouBou75}, $\alpha$ and $\beta$
\begin{align}\label{eq:std_form}
a_{fi} & =
\alpha \, (\hat{\v{e}}\cdot\hat{\v{\varepsilon}} ) \delta_{F_fF_i}\delta_{M_fM_i}\nonumber\\
 & + i \beta \, (\hat{\v{e}} \times \hat{\v{\varepsilon}}  ) \cdot{\bra{F_f M_{f}}\v{\sigma}\ket{F_i M_{i}}} \, .
\end{align}
These two terms map into the $Q=0$ and $Q=1$ contributions in Eq.~{\eqref{eq:st_amp_exp}}, respectively. In other words,
\begin{align} \label{Eq:Tran-pol-2nd-order-tensorial}
    a_{fi}&=-\sqrt{3[F_f]}w_0( \hat{\v{\varepsilon}},\hat{\v{e}})\alpha\nonumber\\
    &-\sqrt{2}\brad{F_f}{\sigma}\ketd{F_i}w_1( \hat{\v{\varepsilon}},\hat{\v{e}}) \beta\,,
%&+\brad{f}\{I\otimes{I}\}^{(2)}\ketd{i}w_2( \hat{\v{\varepsilon}},\hat{\v{e}})\gamma\,.
\end{align}
where, for the $S_{1/2}$ states, the reduced matrix element $\brad{F_f}{\sigma}\ketd{F_i}$ in the hyperfine basis~\eqref{eq:hyperfine-basis} is given by
\begin{align}\label{eq:fsigi}
    \brad{F_f}{\sigma}\ketd{F_i}&=\sqrt{6}(-1)^{I+F_i-1/2}\nonumber\\
    &\times\sqrt{[F_f,F_i]}\sixj{1/2}{F_f}{I}{F_i}{1/2}{1}\,,
%\brad{f}\{I\otimes{I}\}^{(2)}\ketd{i}&=(-1)^{2F_i-F_f+I-J_f}\sqrt{5}\,{[F_f,F_i]}^{1/2} \label{eq:fIIi} \times\\
%&I(I+1)(2I+1)\sixj{1}{1}{2}{I}{I}{I} \delta_{J_i,J_f}\,, \nonumber
\end{align}
where we have used $\brad{S=1/2}{\sigma}\ketd{S=1/2}={\sqrt{6}}$.

In accordance with Eq.~(\ref{eq:afq_exp}), one may then write
\begin{subequations}
    \begin{align}
        \alpha=&-\frac{\brad{f}a[f]^{(0)}\ketd{i}+\brad{f}a[i]^{(0)}\ketd{i}}{\sqrt{3(2F_f+1)}}\,,\\
        \beta=&-\frac{\brad{f}a[f]^{(1)}\ketd{i}-\brad{f}a[i]^{(1)}\ketd{i}}{\sqrt{2}\brad{F_f}{\sigma}\ketd{F_i}}\,,
    \end{align}
\end{subequations}
or, as explicit sums over intermediate states,
\begin{subequations}\label{eq:sec_order_alpha_beta}
\begin{align}
\alpha&=\delta_{J_fJ_i}\sqrt{\frac{1}{6}}\sum\limits_{n_{a}J_a}\frac{(-1)^{J_a-J_i}}{\sqrt{3(2J_i+1)}}\nonumber\\
&\times\brad{n_fJ_f}D\ketd{n_a{J_a}}\brad{n_{a}{J_a}}D\ketd{n_iJ_i}\nonumber\\&\times\left(\frac{1}{E_{n_fJ_f}-E_{{n_aJ_a}}}+\frac{1}{E_{n_iJ_i}-E_{{n_aJ_a}}}\right)\,,\label{eq:sec_order_alpha}\\
\beta
&=-\frac{1}{2}\sum\limits_{n_aJ_a}\sixj{1}{J_i}{J_f}{J_a}{1}{1}\nonumber\\
&\times\brad{n_fJ_f}D\ketd{n_aJ_a}\brad{n_aJ_a}D\ketd{n_iJ_i}\nonumber\\&\times\left(\frac{1}{E_{n_fJ_f}-E_{n_aJ_a}}-\frac{1}{E_{n_iJ_i}-E_{n_aJ_a}}\right)\,.\label{eq:sec_order_beta}
\end{align}
\end{subequations}
These equations recover the conventional expressions in the literature,
see, e.g., formulae in Ref.~\cite{BluJohSap92} specialized for the initial and final states of the $S_{1/2}$ character. In this case,  the $E1$ selection rules fix the intermediate states to the $P_{1/2}$ and $P_{3/2}$ angular characters. 

\section{Evaluation of the transition polarizabilities}\label{Sec:Cal-sec-order}

In the last section, we derived the second order transition polarizabilities $\alpha$ and $\beta$. In this section, we present the numerical methods with which these polarizabilities are calculated. Our approach is a blend of relativistic many-body methods of atomic structure and high-precision experimental values for atomic level energies. 

Since the Cs atom has $55$ electrons, its electronic structure is relatively simple: it has a single valence electron outside the [Xe]-like closed-shell core. This simplicity greatly facilitates accounting for many-body effects due to the residual electron-electron interaction (correlation). In what follows, we describe several approximations of increasing complexity through which the correlation contributions to $\alpha$ and $\beta$ are computed.

The lowest-order approximation in the electron-electron interaction is the mean-field Dirac-Hartree-Fock (DHF) method, wherein each electron experiences an ``averaged'' influence from all other electrons (and of course the Coulomb interaction with the nucleus). Within the DHF approach, we use the ``frozen-core'' approximation, where atomic orbitals in the [Xe]-like closed-shell core are computed self-consistently, and the valence orbitals are determined afterward in the resulting  $V^{N-1}$ DHF potential of the core. 

We point out that even at this lowest-order DHF level, the intermediate states involved in calculating $\alpha$ and $\beta$, 
Eqs.~\eqref{eq:sec_order_alpha_beta}, span a countable yet infinite set of bound states and an innumerable set of states in the continuum. Since this Hilbert space is infinitely large, direct numerical summations, while possible, require different numerical implementations for various many-body methods. An elegant way to handle this issue is the $B$-spline approach popularized by the Notre Dame group~\cite{JohBluSap1988-Bsplines,SapJoh1996-BsplineReview, Joh07}. This approach generates a {\em finite} and numerically complete basis set that has been proven useful in evaluating otherwise infinite sums. In this approach, the set of eigenfunctions is a linear combination of $B$-spline functions covering a radial grid extending from the origin to $R_{\mathrm{max}}$, the radius of an artificially imposed spherical cavity. The Notre Dame approach is further refined by employing the dual-kinetic balance $B$-spline basis set~\cite{BelDer08.DKB} which helps mitigate the issue of spurious states and improve the numerical quality of orbitals both near and far away from the nucleus. 

The low-$n$ orbitals from a $B$-spline finite basis set closely resemble those obtained with the conventional finite-difference techniques with a sufficiently large radial grid extent. We refer to these low-$n$ orbitals as ``physical'' states. As $n$ increases, this mapping deteriorates, so higher-$n$ basis orbitals often differ substantially from their finite-difference counterparts; we refer to such states as ``nonphysical'' 
states. 

The value of $n$ separating the physical and nonphysical parts of the pseudospectrum primarily depends on $R_{\rm max}$ and to some extent on the number of basis functions. The dependence on the cavity's radius is easily understood by recalling that low-$n$ orbitals decays exponentially with increasing distance from the nucleus (origin) so they cannot ``know about'' the existence of a cavity of sufficiently large radius. In contrast, high-$n$ orbitals have their maxima at larger distances and therefore are much more susceptible to the cavity's presence. Our $B$-spline basis set contains $N=60$ basis functions of order $k=9$ per partial wave generated in a cavity of radius $R_{\mathrm{max}}=250\,\mathrm{a.u.}$. These parameters are chosen so that the fractional differences in the DHF eigenenergies between the basis set and the finite-difference approach for physical states ($n' \le 12$) are within $0.015\%$. Similarly, the basis-set values of the $E1$ matrix elements involving physical states differ from their finite-difference counterparts by less than $0.1\%$. A detailed discussion of the proper mapping of the finite basis set orbitals to the physical states may be found in Ref.~\cite{tan2023precision}.
 
With the finite basis set, one may further facilitate the numerical evaluations of $\alpha$ and $\beta$ by splitting the summations in Eqs.~{\eqref{eq:sec_order_alpha}} and {\eqref{eq:sec_order_beta}} into the ``core-valence'' (``cv''), ``main'', and ``tail'' contributions 
\begin{subequations}
\begin{align}  
\alpha&=\alpha_{\rm{cv}}+\alpha_{\rm{main}}+\alpha_{\rm{tail}}\,,\\
\beta&=\beta_{\rm{cv}}+\beta_{\rm{main}}+\beta_{\rm{tail}}\,,
\end{align}
\end{subequations}
where the cv terms correspond to summations over $2\le{n_a}\le{5}$, the main terms to summations over $6\le{n_a}\le{12}$, and the tail terms to summations over $13\le{n_a}\le{\infty}$, respectively. The cv term comes from the core particle-hole intermediate states with excitations to the valence orbital blocked by the Pauli exclusion principle~\cite{DerJohSaf99}. The infinity in $13\le{n_a}\le{\infty}$ corresponds to the maximum number of basis set orbitals of a given angular character. We disregard the summation over Dirac negative-energy states, as their contribution in the length-gauge for dipole operators is suppressed by $\alpha_{\rm fs}^4$, where $\alpha_{\rm fs}\approx1/137$ is the fine-structure constant. We have chosen the boundary $n_a=12$ between the main and tail terms with the convention of the earlier work, Ref.~\cite{Toh2019}, in mind. Since we have carefully chosen our finite basis set so that that $B$-spline single-electron orbitals with $n_a\le 12$ coincide with their finite-difference counterparts, the intermediate many-body states $\ket{n_aJ_a}$ in $\alpha_{\rm{main}}$ and $\beta_{\rm{main}}$ map into physical states. 
   
The next-level approximation is the Brueckner orbitals (BO) method which incorporates certain many-body effects beyond the DHF treatment. BOs
qualitatively describe the phenomenon where the valence electron charge causes the atomic core to become polarized, thus inducing a dipole and higher-rank multipolar moments within the core. Consequently, the redistributed charges within the core attract the valence electron. Compared to the DHF approximation, the BO method improves the theory-experiment agreement for valence electron removal energies. In our work, the BO basis set is obtained by rotating the DHF set using the second-order self-energy operator, see Ref.~\cite{tan2023precision} for further details.

A further improvement upon the DHF and BO methods is the random phase approximation (RPA), which is a linear response theory implemented within the mean-field framework~\cite{amusia1990atomic,Johnson-1989-RRPA}. The primary function of RPA is to  account for the screening of externally applied fields by the core electrons. The main advantages of the RPA formalism are that RPA is an all-order method and the RPA transition amplitudes are gauge-independent. For more details about our finite-basis-set implementation of RPA, the reader is referred to Ref.~\cite{tan2023precision}. The RPA(BO) approach incorporates both the core polarization and the core  screening effects. The quality of the RPA(DHF) and RPA(BO) dipole matrix elements is substantially improved over the DHF or BO methods, see, again Ref.~\cite{tan2023precision}. %The quality of this approximation has been studied extensively in the literature (see, e.g., Refs.~\cite{JohLiuSap96,DzuFlaSus84}).

To proceed beyond RPA(DHF) and RPA(BO), we employ the all-order relativistic many-body coupled-cluster (CC) approach, which systematically accounts for correlation contributions at each level of approximation. In our recent work~\cite{tan2023precision}, $E1$ matrix elements between the $6,7S_{1/2}$ and $nP_{1/2,3/2}$ states for $6\leq n\leq 12$ were computed using the CCSDpTvT method. This method incorporates  single (S), double (D), and triple (T) excitation from the reference DHF state~\cite{tan2023precision} in the CC formalism. The ``pTvT'' qualifier in CCSDpTvT  refers to a perturbative treatment of core triples and a full treatment of valence triples. In addition to an accurate treatment of the many-body effects, the CCSDpTvT $E1$ matrix elements values include scaling, dressing, Breit, and QED corrections~\cite{tan2023precision}. These CCSDpTvT values are the most complete theoretical determinations of the $E1$ matrix elements in Cs to date. The results are complete through the fifth order of many-body perturbation theory and include some chains of topologically-similar diagrams to all orders. As such, the CCSDpTvT method is the most theoretically complete applied to correlation effects in Cs so far. Since the finite basis set used in Ref.~\cite{tan2023precision} is identical to that employed in this work, we identify the CCSDpTvT many-body states with $6\leq n\le{12}$ with the physical states and use the CCSDpTvT matrix elements to compute the main contribution to transition polarizabilities.

\section{Numerical results and discussions}\label{Results and error estimates}

We have provided an overview of the numerical approaches employed in our calculations of the transition polarizabilities $\alpha$ and $\beta$. In what follows, we present our numerical results and estimates of uncertainties.

In Table~{\ref{tab:alpha_beta}}, we compile our numerical results for $\alpha$ and $\beta$. In addition to the DHF, BO, RPA(DHF), RPA(BO), and CCSDpTvT results, we list our values obtained from CC calculations of varying complexity. In particular, in the SD approximation, only linear singles and doubles are included, the CCSD approximation additionally incorporates nonlinear effects, the CCSDvT approximation includes full valence triples on top of CCSD, and finally CCSDpTvT(scaled) indicates a CCSDpTvT value rescaled using experimental values for the removal energies. See Ref.~\cite{tan2023precision} for further details. The final values for $\alpha$ and $\beta$ are obtained by adding to the scaled CCSDpTvT values the Breit, QED, and basis extrapolation contributions to the $E1$ matrix elements, as mentioned in Sec.~\ref{Sec:Cal-sec-order}. Note that the different CC approximations only apply to the $E1$ matrix elements in Eqs.~\eqref{eq:sec_order_alpha_beta}. For the energy denominators, we have used the DHF, BO, RPA(DHF), RPA(BO) values in the corresponding approximations and experimental values for all CC approximations. Note also that the CC approximations were only used to compute the main terms, as mentioned in Sec.~\ref{Sec:Cal-sec-order}. The cv and tail terms are only calculated up to RPA(BO), since (i) their contributions are much smaller than those of the main terms, (ii) full CCSDpTvT calculations are expensive, and (iii) the disparity between the high-$n$ states in the tail terms and their physical counterparts is significant. The semi-empirical result for $\beta$ is obtained by dividing our theoretically determined result for $\alpha$ by the experimentally measured ratio $\alpha/\beta=9.905(11)$~\cite{ChoWooBen97} (see below for further details on this point). 

\begin{table}[ht!]
    \centering
    \begin{tabular}{ld{6}d{7}}
        \hline\hline
    &\multicolumn{1}{c}{$\alpha$}&\multicolumn{1}{c}{$\beta$} \\
    \hline
    \multicolumn{3}{c}{This work}\\
    DHF                                                   & -348.50      & 29.278     \\
    BO                                                    & -339.53      & 26.483     \\ 
    RPA(DHF)                                              & -276.17      & 29.318     \\
    RPA(BO)                                               & -273.68      & 26.364     \\
    SD                                                    & -272.73      & 26.934     \\
    CCSD                                                  & -279.55      & 27.214     \\
    CCSDvT                                                & -266.04      & 27.370     \\
    CCSDpTvT                                              & -265.85      & 27.324     \\
    CCSDpTvT(scaled)                                      & -266.23      & 27.227     \\
    \hline
    Final                                                 & -266.31(23)  & 27.023(114)\\
    \textbf{Semi-empirical} $\boldsymbol{(\alpha/\beta)}$ &              & \textbf{26}.\textbf{887(38)} \\
    \hline
    \multicolumn{3}{c}{Other works}\\
    Sah20~\cite{sahoo20}       (Sum over states $\alpha$) & -268.65(27)  & 27.12(4)   \\
    Toh19~\cite{Toh2019}       (Sum over states $\alpha$) & -268.81(30)  & 27.139(42) \\
    Dzu02~\cite{Dzuba2002}     (Sum over states $\alpha$) &              & 27.15(11)  \\
    Vas02~\cite{Vas2002}       (Sum over states $\alpha$) & -269.7(1.1)  & 27.22(11)  \\
    Dzu00~\cite{DzuFla00}      ($M1$ calculation)         &              & 26.957(51) \\
    Ben99~\cite{Bennett1999a}  ($M1/\beta$ experiment)    &              & 27.024(80) \\
    Saf99~\cite{Safronova1999} (Sum over states $\alpha$) & -268.6(2.2)  & 27.11(22)  \\
    Saf99~\cite{Safronova1999} (Sum over states $\beta$)  &              & 27.16      \\
    Dzu97~\cite{Dzuba1997}     (Sum over states $\alpha$) & -269.0(1.3)  & 27.15(13)  \\
    Blu92~\cite{BluJohSap92}   (Sum over states $\beta$)  &              & 27.0(2)    \\
    \hline\hline
    \end{tabular}
    \caption{Numerical results for the  scalar and vector $6S_{1/2}\rightarrow{7S_{1/2}}$ transition polarizabilities in ${}^{133}$Cs in the Dirac-Hartree-Fock (DHF) approximation, the Brueckner orbitals (BO) approximation, the random-phase  approximation (RPA) implemented on a DHF basis set, RPA implemented on a BO basis set, the coupled-cluster (CC) approximation with only linear singles and doubles (SD), the CC approximation with nonlinear treatment singles and doubles (CCSD), the CC approximation with linear and nonlinear singles and doubles and valence triples (CCSDvT), the CC approximation with linear and nonlinear singles and doubles, perturbative core triples, and valence triples (CCSDpTvT). The CCSDpTvT(scaled) result is obtained by using experimental results for removal energies to rescale single and double amplitudes. The final result is obtained by adding to CCSDpTvT(scaled) the Breit, QED, and basis extrapolation contributions. The semi-empirical result for $\beta$ (in bold) is obtained by combining the theoretically determined result for $\alpha$ with the experimentally measured ratio $\alpha/\beta=9.905(11)$~\cite{ChoWooBen97}. All values are given in atomic units.}
    \label{tab:alpha_beta}
\end{table}

As shown in Table~{\ref{tab:alpha_beta}}, the BO correction has a larger impact on $\beta$ than on $\alpha$, differing by 9.6\% from the DHF value of $\beta$, while being only 2.6\% away from the DHF value for $\alpha$. In contrast, the RPA contribution appears to be much more important for $\alpha$ than $\beta$, with the RPA(DHF) and RPA(BO) values for $\alpha$ differing from the DHF and BO values by around 20\%, whereas the RPA(DHF) and RPA(BO) values for $\beta$ are only 0.1--0.5\% away from the corresponding DHF and BO values. SD shifts $\alpha$ by 0.35\% away from its RPA(BO) value while the SD change for $\beta$ is at 2.2\%. CCSD moves the SD value for $\alpha$ by 2.5\% and that for $\beta$ by 1\%. Adding valence triples amounts to a 4.9\% shifts for $\alpha$ and a 0.6\% shift for $\beta$ while perturbative core triples give rise to a 0.07\% shift for $\alpha$ and a 0.2\% shift for $\beta$. Semi-empirical scaling of removal energies changes $\alpha$ by 0.2\% and $\beta$ by 0.4\%, while Breit, QED, and basis extrapolation corrections are at the level of 0.03\% for $\alpha$ and 0.8\% for $\beta$. 

\begin{figure}
    \centering
    \includegraphics[width=\columnwidth]{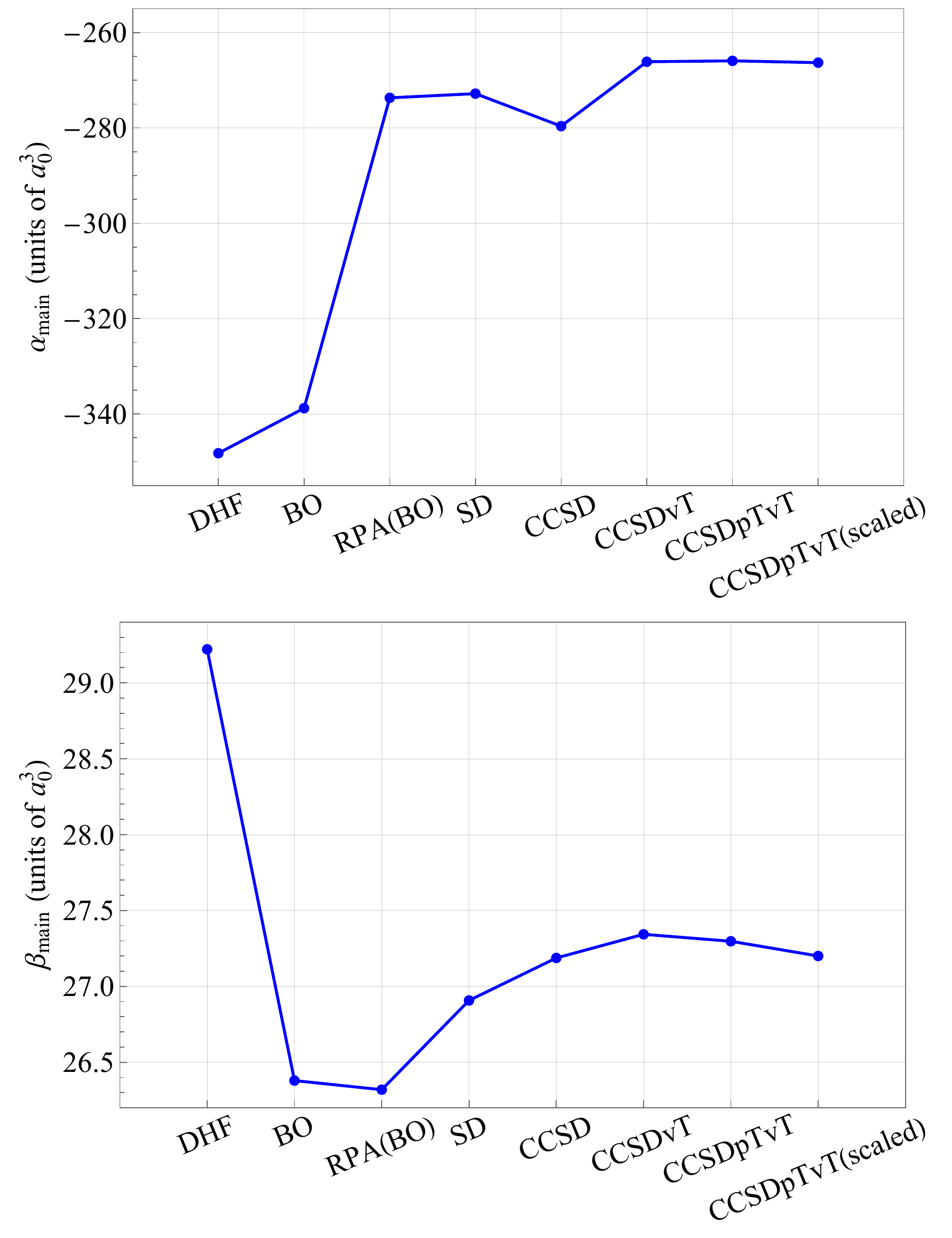}
    \caption{Convergence patterns for the main contributions $\alpha_{\rm main}$ and $\beta_{\rm main}$ to the second order scalar atomic polarizabilities with increasing complexity of the approximations for electron correlation effects.}\label{fig:convpattalphamain}
\end{figure}

In Table~\ref{tab:4approach}, the cv, main, and tail contributions to $\alpha$ and $\beta$ in different approximations are presented explicitly. This allows us to determine the central values for our computations and estimate our uncertainties. The uncertainties in $\alpha_{\rm main}$ and $\beta_{\rm main}$ may be estimated by considering the convergence patterns of these terms across various approximations. Indeed, Fig.~\ref{fig:convpattalphamain} shows the diminishing of contributions from terms of higher and higher order in many-body perturbation theory: the RPA contributions are large, the additional effects of nonlinear core singles and doubles and valence triples, although significantly smaller, are still substantial, whereas additional core triples and scaling effects are generally small. As a result of this observation, we estimate our uncertainty $\sigma_{CC}$ as half the difference between the CCSDpTvT and scaled CCSDpTvT values. This uncertainty represents missing contributions from higher-order CC diagrams. The uncertainty $\sigma_{\rm Breit+QED+basis}$ from the Breit, QED, and basis extrapolation contributions are assumed, conservatively, to be half the difference between the final and scaled CCSDpTvT values. The total uncertainties $\sigma_{\rm main}$ in $\alpha_{\rm main}$ and $\beta_{\rm main}$ are obtained by adding $\sigma_{CC}$ and $\sigma_{\rm Breit+QED+basis}$ in quadrature.

The contributions and uncertainties of the tail terms may be estimated by considering how much the DHF, BO, RPA(DHF), and RPA(BO) values for $\alpha_{\rm main}$ and $\beta_{\rm main}$ differ from the final CCSDpTvT results of these main terms. We observe that the RPA(DHF) and RPA(BO) approximations generally give better agreement with the final values, as to be expected since RPA is known to be responsible for a large portion of the electron correlation effects. As a result, we assume that the contributions from the tail terms are the average of the corresponding RPA(DHF) and RPA(BO) values, and that the uncertainties $\sigma_{\rm tail}$ are half of the corresponding RPA(DHF) and RPA(BO) differences. 

Finally, since the cv terms are the same in the RPA(DHF) and RPA(BO) approaches, we take these values as our estimates for $\alpha_{\rm cv}$ and $\beta_{\rm cv}$. The uncertainties $\sigma_{\rm cv}$ in these contributions are assumed to be half the corresponding BO and RPA(BO) differences. The total uncertainties in our evaluations of $\alpha$ and $\beta$ are obtained by adding $\sigma_{\rm cv}$, $\sigma_{\rm main}$, and $\sigma_{\rm tail}$ in quadrature.

\begin{table}[ht!]
	\centering
	\begin{tabular}{ld{2}d{2}d{2}d{3}d{3}d{3}} 
		\hline\hline
		& \multicolumn{1}{c}{$\alpha_{\rm main}$} & \multicolumn{1}{c}{$\alpha_{\rm cv}$} & \multicolumn{1}{c}{$\alpha_{\rm tail}$} & \multicolumn{1}{c}{$\beta_{\rm main}$} & \multicolumn{1}{c}{$\beta_{\rm cv}$} & \multicolumn{1}{c}{$\beta_{\rm tail}$} \\
		\hline
        DHF              & -348.24 & 0.20 & -0.46 & 29.221 & 0.001 & 0.056 \\ 
        BO               & -338.80 & 0.21 & -0.94 & 26.379 & 0.002 & 0.103 \\ 
        RPA(DHF)         & -276.33 & 0.40 & -0.24 & 29.309 & 0.003 & 0.006 \\ 
        RPA(BO)          & -273.69 & 0.40 & -0.39 & 26.319 & 0.003 & 0.042 \\ 
        SD               & -272.81 &      &       & 26.907 &       &       \\ 
        CCSD             & -279.63 &      &       & 27.187 &       &       \\ 
        CCSDvT           & -266.12 &      &       & 27.343 &       &       \\ 
        CCSDpTvT         & -265.93 &      &       & 27.297 &       &       \\ 
        CCSDpTvT(scaled) & -266.31 &      &       & 27.200 &       &       \\ 
        \hline
        Final            & -266.39 & 0.40 & -0.32 & 26.996 & 0.003 & 0.024 \\ 
        Uncertainty      &    0.20 & 0.10 & 0.09  &  0.113 & 0.001 & 0.018 \\
		\hline\hline
	\end{tabular}
	\caption{The behaviors of the core-valence, main, and tail contributions to $\alpha$ and $\beta$ across several approximations employed in this work. See the caption of Table~\ref{tab:alpha_beta} for an explanation of the notation. All values are given in atomic units.}\label{tab:4approach}
\end{table}

\begin{table*}[!ht]
    \centering
    \begin{tabular}{lcd{6}d{7}d{7}d{7}} 
    \hline\hline
    \multicolumn{1}{c}{} &
    \multicolumn{1}{c}{$n_a$} &
    \multicolumn{1}{c}{Toh19~\cite{Toh2019}} &
    \multicolumn{1}{c}{SD} &
    \multicolumn{1}{c}{Final} &
    \multicolumn{1}{c}{Difference}
    \\ 
    \hline
    \multirow{7}{*}{$n_aP_{1/2}$}
    &6 & -32.54    &            & -32.44     & 9.94[-2] \\ 
    &7 & -37.35    &            & -36.84     & 5.04[-1] \\
    &8 & -5.46[-2] & -5.500[-2] & -4.551[-1] & 8.92[-2] \\
    &9 & -7.99[-2] & -7.824[-2] & -5.611[-2] & 2.41[-2] \\
    &10& -2.30[-2] & -2.239[-2] & -1.374[-2] & 9.48[-3] \\
    &11& -9.31[-3] & -9.036[-3] & -4.839[-3] & 4.38[-3] \\
    &12& -4.61[-3] & -4.472[-3] & -2.007[-3] & 2.81[-3] \\
    \hline 
    \multirow{7}{*}{$n_aP_{3/2}$}
    &6 & -92.93    &            & -92.68     & 2.55[-1] \\ 
    &7 & -102.1    &            & -101.1     & 1.00     \\ 
    &8 & -2.43     & -2.461     & -2.215     & 2.13[-1] \\
    &9 & -4.69[-1] & -4.685[-1] & -4.042[-1] & 6.51[-2] \\
    &10& -1.65[-1] & -1.650[-1] & -1.372[-1] & 2.81[-2] \\
    &11& -7.79[-2] & -7.774[-2] & -6.375[-2] & 1.41[-2] \\
    &12& -4.34[-2] & -4.329[-2] & -3.489[-2] & 8.53[-3] \\
    \hline
    Main (6, 7)  & & -264.86 &         & -263.00 &  1.86 \\ 
    Main (8--12) & & -3.847  &         & -3.387  &  0.46 \\
    Core-valence & &  0.2    &         &  0.40   &  0.20 \\
    Tail         & & -0.30   &         & -0.32   & -0.02 \\
    \hline
    Total        & & -268.81 &         & -266.31 &  2.50 \\
    \hline\hline
    \end{tabular}
    \caption{Comparison of individual contributions to the scalar transition polarizability $\alpha$ from intermediate states $n_aP_J$ with $n_a=6,\ldots,12$, as well as core-valence and tail terms, as computed by using the matrix elements provided by Ref.~\cite{Toh2019} and by us. The notation $x[y]$ stands for $x\times 10^y$. See the caption of Table~\ref{tab:alpha_beta} for an explanation of other notations. All values are given in atomic units.}\label{tab:compare-CCSD-CCSDpTvT}
\end{table*}

In Table~\ref{tab:compare-CCSD-CCSDpTvT}, the main term of $\alpha$ is further broken down into contributions from intermediate states $n_aP_{J}$ with different $n_a$. This facilitates a detailed comparison between the result of this work and that of Ref.~\cite{Toh2019}. We first remind the reader that Ref.~\cite{Toh2019} estimated the contributions from $6,7P_{J}$ by using experimental values for the $E1$ matrix elements between $6,7S_{1/2}$ and these $P$ states. In contrast, we estimate the contributions from $6,7P_{J}$ by using theoretical CCSDpTvT values for the matrix elements from Ref.~\cite{tan2023precision}. For $6,7P_{1/2}$, the two approaches agree quite well, reflecting the fact that the theoretical CCSDpTvT matrix elements $\brad{6,7S_{1/2}}D\ketd{6,7P_{1/2}}$ from Ref.~\cite{tan2023precision} are in good agreement with experiments. On the other hand, our estimates for the contributions from $6,7P_{3/2}$ disagree quite substantially with those from Ref.~\cite{Toh2019}, due to tensions between theoretical values for $\brad{6,7S_{1/2}}D\ketd{6,7P_{3/2}}$ from Ref.~\cite{tan2023precision} and experimental results.

Another noticeable feature of Table~\ref{tab:compare-CCSD-CCSDpTvT} is the significant difference between our values for the contributions from $n_aP_J$ with $n_a=8,\ldots,12$ and those of Ref.~\cite{Toh2019}. This discrepancy is due to the fact that Ref.~\cite{Toh2019} used for matrix elements $\brad{6,7S_{1/2}}D\ketd{n_aP_{J}}$ ($n_a=8,\ldots,12$) theoretical values from Ref.~\cite{Safronova2016}, which computed them in the SDpT approximation. In Table~\ref{tab:compare-CCSD-CCSDpTvT}, we present our SD values for the contributions with $n_a=8,\ldots,12$. One observes that our SD values generally agree with those used by Ref.~\cite{Toh2019}, with the small deviations coming from the pT contributions and the fact that Ref.~\cite{Safronova2016} used a different basis from ours, with less accurate mapping to the physical states. 

We next point out that Ref.~\cite{Toh2019} estimated the tail contribution by first computing $\alpha_{\rm tail}$ in the DHF approximation, then rescaling this DHF result based on the fact that the DHF values for contributions from $n_a=8,\ldots,12$ are $\sim$ 30\% higher than the more accurate SDpT values. In this work, we adopt a slightly different method mentioned earlier, where we estimate $\alpha_{\rm tail}$ by averaging the corresponding RPA(DHF) and RPA(BO) values. This approach stems from the observation that for individual contributions to $\alpha_{\rm main}$ from $n_a=8,\ldots,12$, the average of our RPA(DHF) and RPA(BO) values agree well with the final CCSDpTvT results. Reassuringly, our final value for $\alpha_{\rm tail}$ is in good agreement with that of Ref.~\cite{Toh2019}. We note also that while our DHF value for $\alpha_{\rm cv}$ agrees with that of Ref.~\cite{Toh2019}, we choose to estimate this term using the RPA(DHF) and RPA(BO) methods, since these are more complete theoretical treatments. 

From Table~\ref{tab:compare-CCSD-CCSDpTvT}, the origin of the difference between our estimate for $\alpha$ and that of Ref.~\cite{Toh2019} is also clear. Out of the total disagreement of 2.50 a.u., 1.86 (74\%) comes from the disagreement between experimental and theoretical values for $\bra{6,7S_{1/2}}D\ketd{6,7P_J}$, 0.46 (18\%) originates from our use of the CCSDpTvT instead of SDpT values for the main contributions with $n_a=8,\ldots,12$, and the remaining 0.20 (8\%) comes from the cv contribution.

We close by noting that, as may be observed from the lower panel of Fig.~\ref{fig:convpattalphamain}, the computation of $\beta_{\rm main}$ does not converge  as well with increasingly complex approximations as that for $\alpha_{\rm main}$ (upper panel of Fig.~\ref{fig:convpattalphamain}). Indeed, whereas the final uncertainty in $\alpha_{\rm main}$ is at 0.075\%, the final uncertainty in $\beta_{\rm main}$ is about six times worse, at 0.42\%. This may be understood by noting that in $\beta$, contributions from the $nP_{3/2}$ intermediate states add with an opposite sign to those from $nP_{1/2}$, whereas in $\alpha$, contributions from $nP_{1/2}$ and $nP_{3/2}$ add with the same sign, due to the prefactor $(-1)^{J_a-J_i}$ in Eq.~\eqref{eq:sec_order_alpha}. Since the $nP_{1/2}$ and $nP_{3/2}$ states are degenerate in the nonrelativistic limit, $\beta$ is nonzero solely due to relativistic effects and is thus suppressed compared to $\alpha$. This may also be understood from the observation that the matrix element of a rank-1 tensor (the vector polarizability) between the $L=0$ states (the $S$ states in the nonrelativistic limit) vanishes due to the angular selection rules, while the same matrix element between the $S_{1/2}$ ($J=1/2$) states does not. The cancellation of terms and the resulting suppression of $\beta$ render SoS computations of the vector polarizability less reliable than those for $\alpha$. An improved evaluation of $\beta$ involves, as in previous works, combining our theoretically determined value of $\alpha=-266.30(21)$ with the experimentally measured ratio~\cite{ChoWooBen97} $\alpha/\beta=9.905(11)$ to obtain $\beta=26.887(38)$. This semi-empirical value (in bold) for $\beta$ is also presented in Table~\ref{tab:alpha_beta}. It differs from the value of $\beta=27.139(42)$ of Ref.~\cite{Toh2019} by 0.94\% or 4.4$\sigma$ while is only 0.26\% or 1.1$\sigma$ away from the $M1/\beta$ value of $\beta=26.957(51)$ of Ref.~\cite{DzuFla00}. A comparison between our new value for $\beta$ with previous results is presented in Fig.~\ref{Fig:beta}. We conclude that our determination of $\beta$ brings the two alternative approaches ($\alpha/\beta$ and $M1/\beta$) into an essential agreement.

Finally, a weighted average of our value for $\beta$ and that of Ref.~\cite{DzuFla00} results in 
$$\beta=26.912(30)\,.$$ 
This is the most accurate determination of the vector transition polarizability in Cs to date. Since these two values (ours and that of Ref.~\cite{DzuFla00}) were obtained using different methods, potential cross-correlation effects are anticipated to be suppressed when taking the weighted average. Note that taking weighted average over all the values in Fig.~\ref{Fig:beta} would be incorrect, as all the values on the left panel are statistically correlated.

\section*{Acknowledgements}
We thank D.\ Elliott for a discussion.
This work was supported in part by the U.S. National Science Foundation grants PHY-1912465 and PHY-2207546, by the Sara Louise Hartman endowed professorship in Physics, and by the Center for Fundamental Physics at Northwestern University. 

\bibliographystyle{apsrev4-2}
\bibliography{Thesis.bib}%,library-apd
\end{document}